\documentclass[twocolumn,prl,%
showpacs,preprintnumbers,amsmath,amssymb,nofootinbib]{revtex4}

\input epsf

\usepackage{graphicx}
\usepackage{dcolumn}
\usepackage{bm}

\newcommand{\bea}{\begin{eqnarray}}
\newcommand{\eea}{\end{eqnarray}}
\newcommand{\nn}{\nonumber \\}
\newcommand{\pa}{\partial}

\newcommand{\m}{\vec{m}}
\newcommand{\n}{\vec{n}}
\newcommand{\vv}{\vec{v}}
\newcommand{\zero}{\vec{0}}
\newcommand{\vt}{\vec{\theta}}
\begin{document}
\preprint{}

\title{Large $N$ Reductions and Holography}

\author{FURUUCHI Kazuyuki}
\affiliation{%
\normalsize \sl \vspace*{2mm}\\
Harish-Chandra Research Institute\\
Chhatnag Road, Jhusi, Allahabad 211 019, India\\
{\tt furuuchi@mri.ernet.in}
}%
%
%

\begin{abstract}
The large $N$ reductions in gauge theories
are identified with dimensional reductions with
homogeneous distribution of the eigenvalues of
the gauge field,
and it is used to identify
the corresponding
closed string
descriptions 
in the Maldacena duality.
When one does not take the zero-radii limit, 
the large $N$ reductions are naturally 
extended to
the equivalences between the gauge theories
and the ``generalized" reduced models,
which naturally contain the
notion of T-dual equivalence.
In the dual gravitational description,
T-duality relates
two type IIB supergravity solutions,
the near horizon geometry
of D3-branes, and
the near horizon geometry of 
D-instantons 
densely and homogeneously distributing 
on the dual torus.
This is the 
holographic description
of the 
generalized large $N$ reductions.
A new technique for calculating
correlation functions of local
gauge invariant single trace operators 
from the reduced models is also given.
\end{abstract}

\pacs{11.15.Pg; 11.25.Sq; 11.25.Tq; 11.25.Uv}
\maketitle

\section{Introduction}
The large $N$ limit
of gauge theories leads
to a drastic reduction
of dynamical
degrees of freedom \cite{EK}.
The quantities
in a gauge theory
in $D$ dimension 
can be calculated  
from a much simpler 
reduced model, which 
is obtained by dropping off 
the space-time dependence 
of the original gauge theory.
The crucial condition for
this large $N$ reduction
to take place,
in the case of $SU(N)$
gauge group,
is 
a homogeneous 
distribution
of the eigenvalues of gauge fields,
which preserves the $(Z_N)^D$ symmetry.
This is essentially because
the homogeneous 
distribution
generates space-time momentum
from the gauge group
\cite{BHN,Prsi,DW,GK}.

On the otherhand, the
celebrated Maldacena's
duality conjecture
\cite{malda}
states that
the large $N$ gauge theories
have dual descriptions 
in terms of closed strings
in higher dimensions,
concretely realizing
the large $N$ gauge theory--closed string duality
\cite{tHooft}
and
holography \cite{holo} at the same time.
It is interesting to ask 
how 
the large $N$ reductions are realized in the
dual closed string theory 
via the Maldacena duality.
Recently in the Gopakumar's program
towards a precise formulation of
the large $N$
gauge theory--closed string duality \cite{Gprogram},
I studied 
't Hooft-Feynman diagrams of
correlation functions in gauge theories
compactified on a thermal circle,
to read off the corresponding dual geometries
\cite{mine}.
It was mentioned that 
the technique used for calculating
the thermal correlation functions
was reminiscent to that appeared in
the large $N$ reductions. 
However, this aspect
was not investigated in depth there.
In the present article, 
I clarify
its 
relation to the large $N$ reductions,
and its relevance for finding the
dual holographic realization
in the Maldacena duality.
Some aspects of the large $N$ reductions
will be shown to have simple explanations 
in the dual 
closed string description.

One of the motivations for this study is that
the reduced models are convenient for putting on computers,
and therefore clarifying
the holographic dual description of 
the large $N$ reductions
will lead to the test of the Maldacena duality
by computer simulations.

Another main motivation is that
this gives a concrete way to obtain 
a closed
string theory from the matrix model of
M-theory or the type IIB matrix model,
via the well studied Maldacena duality.

\section{Large $N$ reductions in Maldacena duality}%
In this section, I first review
and extend the argument of \cite{mine}
for how to probe the dual geometry
of the $(Z_N)^D$ symmetric phase
by the correlation functions 
in gauge theories
compactified on a $D$ dimensional torus.
Then, the large $N$ reductions
of the gauge theories are obtained
as a limit where the size of the torus
are taken to zero.
The issue of stability
of the homogeneous 
distribution
will be discussed 
with the comparison with
the stability of the corresponding 
dual geometries.

As an example I take $D=4$ case,
where the boundary description
is naturally identified with some 
$SU(N)$ gauge theory.
Throughout this article
I will work in the planar limit \cite{tHooft}.
I study the case where all fields are in
adjoint representation of the
gauge group.\footnote{%
One can also introduce fields in fundamental
representation and repeat 
the arguments similar to the one below,
but baryons may be missed from such arguments
based on Feynman diagrams \cite{Baryons}.
One may still expect from the dual holographic
descriptions similar to what is discussed 
in this article
that the large $N$ reductions still take place.}
When there are fermions, I put
periodic boundary
conditions on them
in all the compactified directions.\footnote{%
In the thermal case,
fermions obey the anti-periodic 
boundary condition in
the Euclidean time direction.
As long as the phase is in the 
$(Z_N)^D$ symmetric
phase the following argument apply, but
whether which phase is realized 
depends also on these boundary conditions.}
This is necessary for obtaining
the reduced model which reproduces the
original gauge theory results.
The crucial condition for the large $N$ reduction
to take place is that the gauge field
takes the configuration
\bea
\label{symm}
A_{\mu} 
=
\frac{1}{R_\mu}
\mbox{diag}(\theta_\mu^1,\cdots,\theta_\mu^N)
\eea
in an appropriate gauge,
where $\theta_\mu^a$
distributing homogeneously between
$[-\frac{1}{2},\frac{1}{2}]$.
The square expectation value
of the Wilson loops
winding around cycles
of $T^4$ are order parameters of the
$(Z_{N})^4$ symmetry.
They vanish in the $(Z_N)^D$ symmetric phase:
$
\label{square}
\langle |W|^2 \rangle = 0
$
\cite{center},
where
\bea
\label{WilsonLoop}
W=
\frac{1}{N}
\mbox{Tr}P \, \exp i
\oint_0^{2\pi R_\mu}
A_\mu dx^{\mu},
\quad \mu=0,\cdots,3.
\eea
Here $R_\mu$ is 
the compactification radius
in $\mu$-th direction
and
$P$ denotes the path ordering.
Whether the above configuration is realized
or not depends on the theory,
here I am interested in a class
of theories where this is the case.
However, see the discussions 
on the Eguchi-Kawai reduction below.

Suppose one calculates
some field theory correlator
$\langle {\cal O}_1(K_{\mu 1})\cdots {\cal O}_n(K_{\mu n})\rangle$
of gauge invariant local trace operators
${\cal O}(K_{\mu j})$, $\mbox{Tr}\Phi^{I_{j1}}\cdots\Phi^{I_{m_jj}}(K_{\mu j})$ for example.
Here, $K_{\mu j}$ is an external momentum of the $j$-th operator
which takes integer values in the unit of
$\frac{1}{R_\mu}$, and $\Phi^I$'s are adjoint scalars.
I take the background gauge
$D_\mu {\cal A}^\mu =
\pa_\mu {\cal A}^\mu + i [A_\mu,{\cal A}^\mu] =0$, with
${\cal A}^\mu$ being the fluctuating quantum
part of the gauge field 
and the background 
configuration $A_\mu$ being (\ref{symm}).
I quantize the theory through the BRS formalism.
Then, the momenta 
$\frac{n_\mu}{R_{\mu}}$ always appear in the 
combination 
$\frac{1}{R_{\mu}}
(n_\mu \delta_{ab} 
+ \theta_\mu^a-\theta_\mu^b)$.
Furthermore, in the planar limit
one can always associate 
a loop momentum
$\frac{n_\mu^{i}}{R_{\mu}}$
($i = 1, \cdots,\ell$ labels the loop momentum)
with an index loop $a_i$,
and they appear in a specific combination
$ 
\label{comb}
\frac{1}{R_{\mu}}
(n_\mu^{i} + \theta_\mu^{a_i})
$ \cite{Prsi,DW,GK,mine}.%
\footnote{The origin of this combination
is the covariant derivative for adjoint fields.
For a planar
diagram with all adjoint field
propagators, the number of index loop is one more
than the number of momentum loops, but one
index sum factors out and just gives an overall factor $N$
\cite{mine}.}
In the large $N$ limit
one can replace the index sums
with the integrations:
\bea
\label{th2P}
\sum_{a_1 \cdots a_\ell}
G(\frac{\theta_\mu^{i}}{R_\mu})
\rightarrow
(N \prod_{\mu=0}^3 R_\mu
\int_{-\frac{1}{2R_\mu}}^{\frac{1}{2R_\mu}}
dP_{\mu {i}} )
G(P_{\mu {i}})
\eea
where $\frac{\theta_\mu^{i}}{R_\mu}$
was replaced with
$P_{\mu i}$ in the $N \rightarrow \infty$ limit.
As one sums over
the gauge indices,
the sums run over the homogenous distribution
of the eigenvalues of the background gauge field.
Thus the sum over the gauge indices
can be replaced by the integration
over the dual torus.
This is the essential mechanism 
for the large $N$ reductions.
The integrand for the correlator
is a function of $P_{\mu i} + \frac{n_{\mu i}}{R_\mu}$.
Hence the correlator has a form
\bea
&&(\prod_{i=1}^\ell \sum_{n_{\mu i}=-\infty}^\infty
\int_{-\frac{1}{2R_\mu}}^{\frac{1}{2R_\mu}} dP_{\mu i})
G(P_{\mu i}+ \frac{n_{\mu i}}{R_\mu}, K_{\mu j}) \nn
&=&
(\prod_{i=1}^\ell
\int_{-\infty}^{\infty} dP_{\mu i})
G(P_{\mu i},K_{\mu j}) .
\label{loop}
\eea
Thus the full internal 
loop momentum integrations
of the {\em uncompactified} theory
had been recovered. 
In other words,
in the large $N$ limit
the functional forms of the
field theory correlators on $T^4$
with the background (\ref{symm})
coinside with those of the
uncompactified theory (with a trivial gauge field
configuration)
to all orders in perturbation theory.
However, notice that
the external momenta $K_{\mu j}$ still take
discrete values.
Therefore when one performs 
Fourier transformation
to the position space, one obtains the
sum over images of the correlation
functions of the uncompactified theory:
\bea
 \label{corr}
\bigl. G(x_j^\mu)\bigr|_{T^4}
=
\bigl.
\sum_{m_j=-\infty}^{\infty} G(x_j^\mu+2\pi m_j R_\mu)
\bigr|_{R^4}.
\eea
The result (\ref{corr}) was recently
obtained in \cite{mine} 
in the context of the Maldacena duality.
It may be worth noting
that the main ingredients 
in the derivation of (\ref{corr})
had appeared
in the old study 
of the large $N$ reductions \cite{Prsi}.
The new viewpoint brought by \cite{mine} was 
its bulk interpretation:
(\ref{corr}) has a simple interpretation
in the corresponding dual geometry.
In the Maldacena duality,
the geometry of the bulk can be probed
by the gauge theory correlators.
Then, (\ref{corr})
means that the dual geometry
probed by the gauge theory Feynman diagrams
of the compactified theory
with the background (\ref{symm})
is the same to that of the uncompactified
theory, except for the periodic identifications
in $T^4$ directions \cite{mine}.
Recall that the result for correlation functions
of composite operators
(\ref{corr}) is not a trivial consequence of
a simple compactification
in the gauge theory side, 
but the configuration
(\ref{symm}) was crucial:
If one sums over images of 
each field's Feynman propagator on $R^4$
to obtain 
the propagator on $T^4$ 
(say $\langle \Phi^I(x_1) \Phi^J(x_2) \rangle$),
which is appropriate for probing
a geometry corresponding to $A_\mu = 0$
background but not the 
homogeneous 
configuration 
(\ref{symm}),
one does not obtain the sum over images
of the correlation functions of 
the composite operators.

Now I identify the large $N$ reduction with
the zero-radii limit $R_\mu \rightarrow 0$,
so that in the 
first line
of (\ref{loop})
the momentum
summation can be truncated
to $n_{\mu i} =0$.
Then, the original $T^4$ momentum summations
drop out, but the gauge index summations
reproduce the $R^4$ momentum integrations.
This is the essence of 
the perturbative ``derivation" of
the holographic dual description
of the large $N$ reduction.\footnote{%
``Derivation"
assuming that the Maldacena duality
is correct for uncompactified theory.
The 
``derivation"
may be extended to the
non-perturbative one by
using the Schwinger-Dyson equation 
\cite{EK}.}

There are two main options for
taking the zero-radii limit,
corresponding to two types of
the reduced models.
The ``quenched" reduced models
\cite{BHN}
are essentially the models where the condition
(\ref{symm}) is put by hand.
This is actually sufficient for 
a purpose of
calculating
quantities
of the {\em uncompactified} original gauge 
theories.
By construction the dual geometry in this case
is the same as that
of the uncompactified theory,
up to 
the periodic identifications
in $T^4$ directions.
To calculate quantities
which is
translationally invariant
along the $T^4$ directions 
from the closed string side, 
one just needs to study
translationally invariant solutions of
classical equation of motions.
The periodic identifications
in $T^4$ directions,
in particular $R_\mu \rightarrow 0$
limit,\footnote{%
The limit is, however, slightly subtle
for conformal field theories where the
small volume limit can be
undone by conformal transformation
(or isometry in dual closed string description).
It will be more appropriate to keep
$R_\mu$ finite in such cases.
This is discussed in the next section.}
do not matter in this case.
This is the closed string dual description
of the quenched large $N$ reduction.
The fact that in
the classical bulk theories
one can truncate the equation of motions
to the holographic radial direction,\footnote{In quantum
theories, even if one is interested
in translationally invariant quantities,
space-time dimensionality
comes in through the loop integrals.}
and the classical limit 
of the closed string theory
corresponds to
the planar limit,
shows a beautiful correspondence between the
two descriptions.
The large $N$ gauge theories may be said to be 
``classical" in this sense.

So far, I have been describing how
the translationally invariant quantities 
can be obtained from the
reduced model, but
the reduced model can also be used to calculate
the quantities which 
depend on space-time coordinates.
This will be explained in the next subsection.

On the otherhand, in the
spirit of the
original reduced model
of Eguchi and Kawai \cite{EK},
the configuration (\ref{symm}) is not
put by hand,
but it must be realized as a
dominant saddle point.
Thus, whether the large $N$ reduction
takes place or not becomes a dynamical issue.
This translates via the Maldacena duality
into the issue of stability of
the geometry dual to the uncompactified
theory upon the zero radii limit of the 
$T^4$ compactification.

The dynamical stability of the 
homogeneous distribution
(\ref{symm}) against
the small volume limit
$R_\mu \rightarrow 0$ 
in gauge theories is a
model dependent problem.
Here I just make a few remarks on 
some
aspects of it.

In the supersymmetric case,
the results of \cite{cmpct}
for $S^1$ compactification
may seem to suggest 
the stability of the configuration (\ref{symm}).
But since here all space-time directions
are compactified, the
quantum fluctuations can be
suppressed only
by the large $N$ effect. 
Therefore
a separate study is actually in order.
Below, I will discuss
a role of fermions
with the periodic boundary conditions,
for the stability
of the configuration (\ref{symm}).

If the gauge theory contains
a massless elementary fermionic field,
the periodic boundary conditions
on it may restrict the topology of
the dual geometry
to be $R_{\geq 0} \times T^4$ \cite{KK}.
This is because if some of the circle
of $T^4$ shrinks to zero
at some distance
in the holographic radial direction in the bulk,
the bulk fermion which couples
to the gauge theory operator
containing the massless fermion
cannot have 
the periodic boundary conditions.\footnote{%
However, this restriction
may not be so strong if one takes into account
other space-time directions in the dual theory. 
See
\cite{Ross} for a recent interesting example
where the circle in the asymptotic boundary
is mixed with
another circle corresponding to an internal symmetry
in field theory side.}
As argued above,
the stability
of the $R_{\geq 0} \times T^4$
topology in the bulk is necessary
for the stability of the configuration (\ref{symm})
in the limit
$R_\mu \rightarrow 0$ 
\cite{mine}.\footnote{%
The bulk topology may also be probed 
by using the classical closed string worldsheet 
as a dual description 
for the Wilson loop expectation values
\cite{Wilson}.
Precisely speaking, what is calculated 
in \cite{Wilson}
is a generalization of the Wilson loop including
adjoint scalars.}
This 
expectation from the closed string side
may heuristically be explained
in the reduced models if one recalls
the procedure taken here
for taking the large $N$ limit.
To see this,
I first analyze
a reduced model with $SU(2)$ gauge group
and with one 
massless adjoint fermion,
to estimate effective potential between two
eigenvalues of the gauge field.
In this case it is possible to integrate out
the fermion \cite{red}, and
it is easy to see that
the presence of the massless
adjoint fermion
introduces a repulsive
potential $\sim - \log L$ for $L \sim 0$,
where
$L$ is a difference between
the two eigenvalues.
One may expect that there
is a similar repulsive force between
eigenvalues also in the $SU(N)$ reduced model.
Then, recall that to obtain the reduced models
from the gauge theories,
I took $N \rightarrow \infty$
before taking the $R_\mu \rightarrow 0$
limit.
To implement this condition
starting from the reduced models, 
one should restrict eigenvalues of the reduced
gauge fields between
$-\frac{1}{2R_\mu}$ and $\frac{1}{2R_\mu}$.%
\footnote{This is a gauge invariant condition
for the reduced models.
The mutually commuting configuration (\ref{symm}) 
should emerge dynamically.}
It is like putting particles
with short distant repulsive forces
dense enough in a finite volume,
so that the resulting distribution
becomes uniform, i.e. the configuration
(\ref{symm}) is realized.
On the otherhand, 
adding a mass term to the fermion
weaken the repulsive force and
it disappears if the mass
is sufficiently large.
It is natural because
if the mass
is taken large enough,
the fermion will eventually
decouple from
the system.
It suggests that
in the corresponding classical solution
of the dual closed string descriptions,
fermionic fields which couple
to a gauge invariant 
fermionic operator with that
massive fermion
are excluded from a 
region in the space-time
corresponding to the scale
lower
than the fermion mass scale,
and they do not restrict the
topology there.
The complete exclusion of
fermionic field from some region may
require a singular geometry
in the supergravity description.
The arguments given here 
are heuristic and deserves further study.

If one introduces a bosonic adjoint
field $\Phi$ to a gauge theory instead
of the fermion,
it means introducing
another space dimension
in the dual closed string side.
Here I study the simple situation
where $\Phi =0$ vacuum is realized
in the gauge theory.
To construct a corresponding
reduced model,
one should take
the diagonal components of $\Phi$ to be zero by hand,
much in the same spirit as in the quenched reduced
models, but for the opposite type of
configuration.\footnote{One may also
try to show that this configuration is
dynamically preferred 
in the reduced model \cite{red}.}
Then, a calculation similar to the above
shows, in $SU(2)$ case,
that the bosonic adjoint field
does not change the leading
repulsive potential from the
fermionic field.

For purely bosonic theories,
in the closed string side
the $AdS$ soliton \cite{adssoliton}
which is a possible vacuum state at finite
$R_\mu$
already partially breaks (\ref{symm}),
and one must also take into account 
the possibilities of various
phase transitions, like
to black holes,
black strings \cite{GL,T}
and so on, 
which can trigger instability
of the configuration (\ref{symm})
upon taking the $R_\mu \rightarrow 0$ limit.
As I mentioned earlier,
the configuration (\ref{symm}) is crucial
for the large $N$ reduction.
The instability of the geometries dual to
the uncompactified theories
upon compactification means that 
the Eguchi-Kawai reduction does not 
take place in those cases.

\subsection{Correlators of local gauge invariant 
single trace operators
from reduced models}
As has been described in the previous
subsections,
the large $N$ reduction is not merely a
simple dimensional reduction
in the gauge theory side, but the 
configuration (\ref{symm}) was crucial.
The loop momentum integrations in the original
gauge theory are
recovered from the gauge index sums
in the reduced model.
One can also calculate correlation functions
in gauge theories
which depend on external momenta
from reduced models,
provided that the configuration (\ref{symm})
is realized.
This is essentially because the gauge indices in the
reduced models play the role of space-time momenta
in the original gauge theory.
In the Maldacena duality,
the correlation function of single 
trace operators
are important because
they correspond to closed string amplitudes
in the dual theory.
However, since trace operators
do not have
un-contracted gauge indices,
additional techniques are required to
calculate 
correlation functions
which depend on external momenta.\footnote{%
If there are un-contracted gauge indices,
these can be straightforwardly regarded
as the external momenta in the large $N$ reductions
\cite{Prsi,DW,GK}.}
In this subsection, I explain
how to calculate such correlation functions
from the reduced models.

\begin{figure}
\begin{center}
 \leavevmode
 \epsfxsize=40mm
 \epsfbox{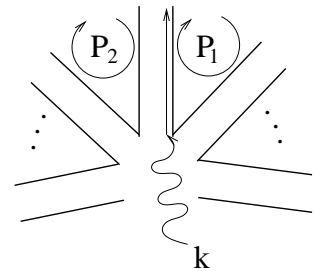}
\end{center}
\caption{The reduced model Feynman diagram 
for a correlator with incoming
$\mbox{Tr}^{\Delta k} \Phi^{I_1}\cdots\Phi^{I_q}$
in the 't Hooft's double line notations.
The figure expresses the assignment of 
the incoming momenta
$k$ to the $P_1$ index loop:
$(P_1+k)-P_2$.}
\label{ch1}
\end{figure}

\begin{figure}
\begin{center}
 \leavevmode
 \epsfxsize=40mm
 \epsfbox{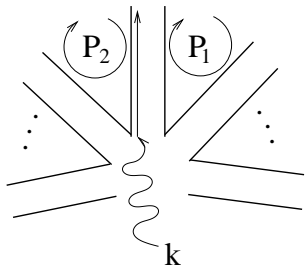}\\
\end{center}
\caption{$P_1-(P_2-k)$.}
\label{ch2}
\end{figure}

\begin{figure}
\begin{center}
 \leavevmode
 \epsfxsize=40mm
 \epsfbox{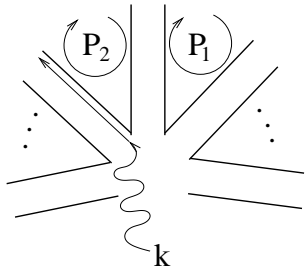}\\
\end{center}
\caption{FIG. \ref{ch2} $\rightarrow$ 
Shift of the integration variable: 
$P_2 \rightarrow P_2 + k$.}
\label{ch3}
\end{figure}

I take an operator
made of adjoint scalars $\Phi^I$
($I$ labels the species of the scalars)
as an example, generalizations to
include fermions or dynamical gauge fields are 
straightforward.
For an operator made of $q$ scalars
$
\mbox{Tr} \Phi^{I_1}\cdots\Phi^{I_q}(k)
$
in the gauge theory,
the corresponding object in the
reduced model is given by
\bea
\label{shifk}
\mbox{Tr}^{\Delta k} \Phi^{I_1}\cdots\Phi^{I_q}
\eea
where I have defined the ``shifted trace"
$\mbox{Tr}^{\Delta k}$
as
\bea
\mbox{Tr}^{\Delta k} AB
=
\sum_{P_1,P_2}
A_{P_1,P_2}B_{P_2,P_1 + k}.
\eea
I have reparametrized the matrix indices
in terms of $P_\mu$, already
taking into account the 
$N \rightarrow \infty$ limit. 
See (\ref{th2P}).
$k_\mu$ is regarded as a shift in the matrix indices.
The point is that the shift
inserts the external momentum $k_\mu$
to the index line connected to 
the latter index of $\Phi^{I_q}$.
The shifted trace does not satisfy
the cyclic property 
$\mbox{Tr} AB = \mbox{Tr} BA$,
so the above mapping from
the gauge theory to the reduced model is
not unique.
However this is not a problem, since
for a planar diagram
these cyclically shifted operators 
all give the same results.
This is because the
index loops and the
incoming momenta
always appear
in the combination
$P_{\mu i_1}-P_{\mu i_2} + k_\mu$,
where $i_1, i_2$ are loop indices.
$k_\mu$ can be assigned to
either $i_1$ or $i_2$,
and the difference 
can be absorbed by a shift of
the loop momentum, which is an integration
variable.
See FIG. \ref{ch1}-\ref{ch3}.%

\section{Finite radii and T-duality}%
\subsection{T-duality in gauge theories}
In the previous section
I identified the large $N$ reductions
with the dimensional reductions
with the non-trivial gauge configuration (\ref{symm}).
However, since
the Maldacena duality is supposed to hold
for any radii,
it is natural to generalize the notion of the
large $N$ reductions to that case.
In this subsection, 
I will explain that the equivalence
between the original gauge theory and the
reduced model still holds for finite radii,
by appropriately generalizing the 
notion of the reduced model 
as a matrix model on a compact space,
along the line of \cite{DT}.
As found in \cite{DT}, 
this naturally leads to the notion of
T-dual equivalence in the matrix model.
The corresponding dual closed string description
of this T-dual equivalence \cite{Kikkawa}
will be presented 
in the subsequent subsection.

\begin{figure}
\begin{center}
 \leavevmode
 \epsfxsize=60mm
 \epsfbox{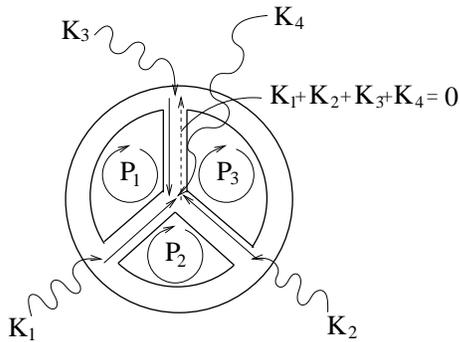}\\
\end{center}
\caption{An example: A four point function
of the operators of the form
$\mbox{Tr}_{SU(N\times M^4)}^{\Delta k, \Delta \m_s}
\Phi^3$ $(s=1,2,3,4)$.
Assigning loop momenta to matrix index loops
automatically taken care of
momentum conservation
at each vertex.
The $\mbox{Tr}_{SU(N\times M^4)}^{\Delta k,\Delta \m_s}
\Phi^3$
adds incoming momentum
$K_s = k_s + \frac{m_s}{R}$
between two index lines.
Note that
$K_1+K_2+K_3+K_4=0$ by the total
momentum conservation
for incoming momenta.
There are different
but equivalent assignments
for between which two index lines
one puts an incoming momentum,
see FIG. \ref{another}.}
\label{income}
\end{figure}

\begin{figure}
\begin{center}
 \leavevmode
 \epsfxsize=60mm
 \epsfbox{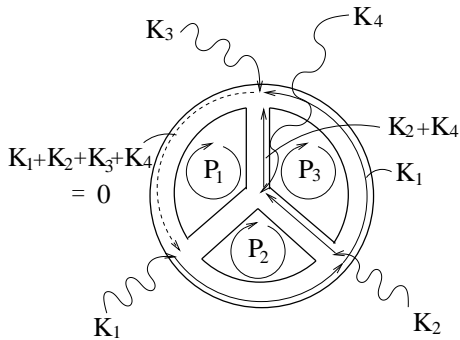}\\
\end{center}
\caption{Another way to calculate the same diagram.}
\label{another}
\end{figure}

For finite radii $R_\mu$,
the gauge theory calculation
has equivalent
T-dual
descriptions in terms
of the matrix model \cite{DT}.
Each eigenvalues are
interpreted as
positions of D-instantons (in the string units) 
in the T-dual language.
The radii of the dual torus $\tilde{T}^4$
are $\frac{\ell_s^2}{R_\mu}$,
where 
$\ell_s$ is the string length.
The summation over $n_\mu$ in (\ref{loop})
corresponds to 
the summation over images of D-instantons
on the dual torus
$\tilde{T}^4$.
To incorporate the images
in the reduced models,
one embeds the $M^4$
$SU(N)$ gauge groups
into the diagonal blocks in
$SU(N\times M^4)$ gauge group,
where $M$ is a positive integer
which will be taken to infinity.
The matrix components of the reduced
fields are subject to
an identification
corresponding to the 
$\tilde{T}^4$
compactification.
The background gauge field configuration
(\ref{symm}) is generalized to
\bea
\label{symmm}
A_{\mu \m\m}
=
\frac{1}{R_\mu}
\mbox{diag}(\theta_\mu^1,\cdots,\theta_\mu^N)
+\frac{m_\mu}{R_\mu}
\eea
where $m_\mu$ is a component
of a four-vector $\m$ which is 
an index for $SU(M^4)$.
The off-diagonal components 
(in terms of $SU(M^4)$)
are zero.
In the matrix model on the torus,
the fields in adjoint representation
satisfy \cite{DT}
\bea
\label{shift}
\Phi_{\vt_1(\m+\vv),\vt_2(\n+\vv)} 
= \Phi_{\vt_1\m,\vt_2\n}
\eea
where
I have labeled the $SU(N)$ gauge group indices
in terms of $\vt$,
and $\m,\n$ are $SU(M^4)$ indices.
$\vv$ is an arbitrary four-vector
with integer entries, which expresses
a parallel shift to an image.
For simplicity, I study 
massless scalar fields $\Phi^I$. 
Generalization
to other fields is
straightforward.
The quadratic term 
of the reduced model is given by
\bea
&&
\frac{1}{M^4}
\mbox{Tr}_{SU(N\times M^4)}
[A_{\mu},\Phi^I][A_\mu,\Phi^I]\nn
&\sim&
\sum_{\m,\vt_1,\vt_2}
\Phi_{\vt_1\m,\vt_2\zero}^I
\left(\frac{1}{R_\mu}
(\theta_{\mu 1}-\theta_{\mu 2} +m_\mu) \right)^2
\Phi_{\vt_2\zero,\vt_1\m}^I\nn
&&
\eea
where use has been made for (\ref{shift}).
In the Maldacena duality, one studies
the coupling of gauge invariant operators
to their sources.
For example, in the gauge theory
the trace of $q$ scalar fields have
the coupling of the form
\bea
\int d^4K
{\cal J}_{I_1\cdots I_q}(-K)
\mbox{Tr}_{SU(N)} 
\Phi^{I_1} \cdots \Phi^{I_q}(K).
\eea
The source ${\cal J}_{I_1\cdots I_q}$ 
is identified with
the boundary value of the corresponding
field in closed string side.
In the reduced models,
the corresponding coupling is given by
\bea
\label{redsrc}
&&(\prod_{\mu}
\int_{-\frac{1}{2R_\mu}}^{\frac{1}{2R_\mu}} dk_\mu
\sum_{m_\mu=-\infty}^\infty )\nn
&&\quad {\cal J}_{I_1\cdots I_q}(-K)
\mbox{Tr}_{SU(N\times M^4)}^{\Delta k,\Delta \m}
\Phi^{I_1} \cdots \Phi^{I_q}
\eea
where $K_\mu = k_\mu + \frac{m_\mu}{R_\mu}$,
and 
I have introduced the
``$\m$-shifted trace" 
$\mbox{Tr}_{SU(M^4)}^{\Delta \m}$
for $SU(M^4)$ indices
defined by
\bea
\label{cyclic}
\mbox{Tr}_{SU(M^4)}^{\Delta \m} AB
&=&
\sum_{\m_1,\m_2}
A_{\m_1,\m_2}
B_{\m_2,(\m_1+\m)} \nn
&=&
\sum_{\m_1,\m_2}
B_{\m_2,\m_1}
A_{\m_1,(\m_2+\m)}.
\eea
The $SU(M^4)$ matrices
$A$ and $B$ satisfy the same condition
as in (\ref{shift}), 
and the
the ``cyclic" property, i.e. 
the second line of (\ref{cyclic}) follows from
that.
The shifted trace for the $SU(N)$ indices $P$
is defined as in (\ref{shifk}). 
One can check
that the reduced model
with the above
source term gives the same result
of the original gauge theory
in the diagrammatic perturbative calculations.
See the example FIG. \ref{income},\ref{another}.
In the dual D-instanton descriptions,
the T-dual of momenta
are winding modes of closed strings,
and the shift $k$ and $\m$ in the trace
$\mbox{Tr}_{SU(N\times M^4)}^{\Delta k, \Delta \m}$
corresponds to a string
stretched between a D-instanton
and another D-instanton shifted by 
$\ell_s^2 (k_\mu + \frac{m_\mu}{R_\mu})$.

Note that usually the sum
over images is not taken in 
the reduced models.
In this sense the
this is a
generalized
of the large $N$ reduction.

\subsection{The dual geometry}
Finally,
I explain in this subsection
that the generalized large $N$
reduction has a simple description
in the dual geometry.\footnote{%
I thank R.~Gopakumar and K.~P.~Yogendran 
for stimulating my thought on T-dual
geometries at the early stage
collaboration in \cite{mine}.}
As a concrete example, I take 
${\cal N}=4$ super Yang-Mills theory
on $T^4$,
which is identified as
a worldvolume theory of D3-branes,
at strong coupling.
At strong coupling, 
supergravity approximation
is valid and
the dual geometry is $AdS_5 \times S^5$
with the periodic identifications in
the $T^4$ directions, and the dilaton is 
constant.
As will be shown below,
this geometry
can be obtained from a
multi D-instanton solution in type IIB supergravity
via T-duality, where D-instantons are
densely and homogeneously
distributing on the dual $\tilde{T}^4$.\footnote{%
The T-dual relation of these geometries
has appeared in \cite{Kallosh}.
The point of this subsection
is to exhibit the parallel
between the dual descriptions.}
The dense homogeneous
distribution of the D-instantons
is identified as 
a holographic dual of the
dense and homogeneous distribution
of the eigenvalues (\ref{symm}).
Thus this is a holographic description
of the equivalence between
the gauge theory and the 
generalized reduced model.

The (Euclideanized) 
metric for the D-instantons in Einstein frame
is flat: $\tilde{g}_{\mu\nu\,E}=\delta_{\mu\nu}$,
$\mu,\nu = 0, \cdots, 9$ \cite{DIS}.
The solution can be obtained by solving
the following equation for dilaton 
$\tilde{\phi}$:
\bea
\pa_\mu \pa^\mu e^{\tilde{\phi}} =0.
\eea
When D-instantons are densely and homogeneously
distributing in the $\tilde{T}^4$ directions,
and overlapping on a point in the transverse six dimensions,
the solution is given by
\bea
e^{\tilde{\phi}_\infty+\tilde{\phi}} 
= 
g_s \left( 1+\frac{c_0 g_s N \ell_s^4}{r^4} \right)
\eea
where $r$ is the radial coordinate
transverse to $\tilde{T}^4$,
$N$ is a number of D-instantons on $\tilde{T}^4$
and $g_s=e^{\tilde{\phi}_\infty}$ 
is the string coupling constant.
$c_0$ is a numerical constant
related to the volume of the unit five-sphere,
I suppress 
such numerical factors hereafter.
In the near horizon limit $r \rightarrow 0$,
the dilaton configuration becomes
\bea
\label{instdilaton}
e^{\tilde{\phi}} = \frac{g_s N \ell_s^4}{r^4}
\eea
and 
I obtain the $AdS_5 \times S^5$ 
metric in the string frame
$\tilde{ds}_{st}^2 =
e^{\tilde{\phi}/2 }\tilde{ds}_{E}^2 $:
\bea
\label{tads}
\tilde{ds}_{st}^2 
=
\frac{\sqrt{g_s N\ell_s^4}}{r^2} 
\left[
dr^2 + r^2 d\Omega_5
+ 
d\tilde{x}_{\scriptscriptstyle /\!\!/}^2
\right] 
\eea
where $\tilde{x}_{\scriptscriptstyle /\!\!/}^\mu$ is
a coordinate on $\tilde{T}^4$ with period 
$2\pi \tilde{R}_\mu$
\footnote{Before taking the near horizon
limit $\prod_{\mu=0}^3 \tilde{R}_\mu = \ell_s^4$
should hold in this solution
so that $N$ coinsides
with the number of D-instantons.
After the near horizon limit
this restriction can be removed 
by the isometry of $AdS_5$.}
and
$d\Omega_5$
is the volume form of the unit five-sphere.
Now I perform T-dual transformation on $\tilde{T}^4$.
The T-dual metric is again $AdS_5 \times S^5$:
\bea
\label{ads}
ds_{st}^2 
= 
\frac{\sqrt{g_s N\ell_s^4}}{r^2} 
\left[
dr^2 + r^2 d\Omega_5
\right] 
+
\frac{r^2}{\sqrt{g_s N\ell_s^4}}
dx_{\scriptscriptstyle /\!\!/}^2
\eea
where $x_{\scriptscriptstyle /\!\!/}^\mu$
is a coordinate on $T^4$
with period $2\pi R_\mu 
= 2\pi \frac{\ell_s^2}{\tilde{R}_\mu}$.
Under the T-duality
the dilaton transforms as \cite{Buscher}
\bea
\label{dilatra}
\phi = \tilde{\phi} 
-
\frac{1}{2}\log \det{}_{\tilde{T}^4} \tilde{g}_{\mu\nu\,st}
= 0.
\eea
Thus one arrives at the $AdS_5\times S^5$
geometry with the constant dilaton
($e^{\phi_\infty}=g_s$), 
as I have claimed.
This is the holographic
description
of the generalized large $N$ reduction
in the previous subsection.
Notice the key role of
the dense and homogeneous
distribution of the D-instantons,
which is dual to the 
dense and homogeneous
of the eigenvalues of the gauge field:
It gives the geometry
which is T-dual to 
the geometry
just obtained by a simple 
$T^4$ identification
of the {\em uncompactified} 
D3-brane near horizon geometry.

Recall that the T-dual relation
in the supergravity classical
solutions can be derived
from closed string worldsheet sigma model
\cite{Kikkawa,Buscher},
whereas the matrix model T-duality
was motivated and ``explained" 
by the open stirng sigma model but was shown purely
within the gauge theoretical language in \cite{DT}.
The validity of these two descriptions
may have an overlap
in the Maldacena's large $N$ 
and the near horizon limit,
as long as the conjecture is 
correct.\footnote{Practically,
one needs to be able to
handle either the stringy corrections 
or
the strongly coupled gauge theory.
Note that although one obtains a smeared solution
from D3-brane solution even for finite $N$,
when the number of the D3-brane
is small the gauge theory description
is not rigorously related to this geometry.
The 't Hooft-Maldacena limit provides the correspondence
between the gauge theory and the closed string theory, 
and 
the large $N$ limit of the homogeneous 
distribution of the eigenvalues of the gauge field,
which is dual to the dense and homogeneous 
distribution of the D-instantons,
provides the effective smearing of the multi-D-instanton
solution.}
Then,
the T-dual equivalence of two geometries
can be interpreted as a holographic
dual description of
the matrix model T-dual equivalence between
the gauge theory and the generalized
reduced model studied in the previous subsection.

\section{Summary and Discussions}
In this article, I 
have presented
the holographic dual descriptions of
the large $N$ reductions 
in the Maldacena duality.
This will be useful for deepening 
the understanding of both sides. 
The equivalence between the reduced model
and the original gauge theory
can be interpreted as a limit of 
the compactification
with the homogeneous 
distribution of the eigenvalues 
of the gauge field.
It was shown how this equivalence 
is reflected in the dual bulk geometry
through the correlation functions
of the local
gauge invariant single trace operators.
Since the 
Maldacena duality holds
even for finite radii, 
it is natural to generalize
the equivalence relation
to that case.
This was achieved by
using the description of the matrix model
on a compact space introduced in \cite{DT}.
This description naturally contains
the notion of T-duality.
I pointed out that
for finite radii the T-dual equivalence
of two supergravity solutions
are the holographic dual description
of the T-dual equivalence between
the gauge theory and the generalized 
reduced model.

The crucial condition for the large $N$
reduction is the 
homogeneous distribution of the eigenvalues of the
gauge field (\ref{symm}). 
In the quenched reduced models 
this condition is forced by hand, whereas
in the Eguchi-Kawai reduction the
stability is a dynamical issue.
The stability of the 
homogeneous distribution
should reflect the stability of the supergravity
solution dual to the uncompactified gauge theory
upon compactification on $T^4$.
I pointed out an interesting possible 
role of fermions obeying the periodic boundary
conditions in the $T^4$ directions.

I also presented a new technique for calculating
position dependent
correlation functions of gauge invariant 
single trace operators in gauge theories from
the reduced models.

Despite the evidences from the past studies,
the Maldacena duality still remains as a conjecture.
The holographic dual of the 
large $N$ reductions established in this
article will
be useful for the quantitative
tests of the Maldacena
duality.
Reduced models are suitable
for studying the non-perturbative effects.
In the Maldacena duality, it is also expected
that the classical closed string descriptions
capture the
non-perturbative effects of the dual gauge theories.
It will be interesting to study further 
how non-perturbative effects in the reduced models
reflect themselves
in the dual closed string descriptions.
The large $N$ reductions 
also provide an advantage 
for computer simulations
\cite{NJ}.
As shown in this
article, the large $N$ reductions
have more direct correspondence with
the Maldacena duality
compared with the lattice 
gauge theory, at least at present.
If a computer simulation
of a reduced model
supports the dynamical stability of
the configuration
(\ref{symm}),
it suggests that
there is a dual closed string solution
which is stable against
the limit 
$R_\mu \rightarrow 0$.
Then one can further calculate
quantities in closed string 
theory using the reduced model.
I hope that
the holographic dual descriptions
of the
large $N$ reductions described
in this article will lead to
the investigation of the Maldacena duality
by computer simulations
of reduced models.

I think relating 
the Matrix model of M-theory \cite{BFSS} or
the IIB matrix model \cite{IIBM}
to the Maldacena duality\footnote{%
I am aware that many researchers have resorted
the idea of relating
the matrix models with Maldacena duality
from different directions.}
via the large $N$ reductions discussed
in this article
is the most
concrete way to study how
closed strings emerge from these models,
especially taking into account
the recent developments in the
understanding of 
the Maldacena duality 
\cite{Gprogram,mine,spin}.

\begin{center}
* * *
\end{center}

I am grateful to my 
colleagues in the present and past institutions
from whom I have learned a lot
about the ingredients appeared in this article.
I am also grateful to my colleagues in HRI
for their continuous warm support.
I sincerely appreciate generous supports 
for our research from 
the people in India.

\end{document}